\documentclass[3p,12 pt,sort&compress, twocolumn]{elsarticle}

\usepackage{color}
\usepackage{amsmath}
\usepackage{subfigure}

\usepackage{setspace}

\usepackage{amsmath}
\usepackage{xfrac}
\usepackage{xcolor}

\usepackage{dcolumn}   % needed for some tables
\usepackage{amsfonts}
\usepackage{booktabs}
\usepackage{siunitx}
\usepackage{multirow}
\usepackage{mathtools}

\usepackage{float}

 \journal{xxxxxx}

\begin{document}

\begin{frontmatter}

\title{Deep Learning and Crystal Plasticity: A Preconditioning Approach for Accurate Orientation Evolution Prediction}					% Activate to display a given date or no 

\author[Queens,NRC]{Peyman Saidi\corref{cor1}\corref{cor2}}
\ead{peyman.saidi@.queensu.ca}
\author[Gent]{Hadi Pirgazi\corref{cor1}}
\author[UNB]{Mehdi Sanjari}
\author[Strathclyde]{Saeed Tamimi}
\author[UNB]{Mohsen Mohammadi}
\author[Queens]{Laurent K. B\'{e}land}
\author[Queens]{Mark R. Daymond}
\author[NRC]{Isaac Tamblyn\corref{cor2}}
\ead{isaac.tamblyn@nrc.ca}

  \cortext[cor1]{P. Saidi and H. Pirgazi are co-first authors }
 \cortext[cor2]{Corresponding Authors:}
 
\address[Queens]{Department of Mechanical and Materials Engineering, Queen's University, Kingston, ON, Canada}
\address[NRC]{National Research Council of Canada, Ottawa, Ontario K1N 5A2, Canada}
\address[Gent]{Department of Electromechanical, Systems and Metal Engineering, Ghent University, Ghent, Belgium}
\address[UNB]{Marine Additive Manufacturing Centre of Excellence (MAMCE), University of New Brunswick, Fredericton, NB, E3B 5A1, Canada}
\address[Strathclyde]{Advanced Forming Research Centre (AFRC), University of Strathclyde, Glasgow, United Kingdom }

\begin{abstract}
Efficient and precise prediction of plasticity by data-driven models relies on appropriate data preparation and a well-designed model. Here we introduce an unsupervised machine learning-based data preparation method to maximize the trainability of crystal orientation evolution data during deformation. For Taylor model crystal plasticity data, the preconditioning procedure improves the test score of an artificial neural network from 0.831 to 0.999, while decreasing the training iterations by an order of magnitude. The efficacy of the approach was further improved with a recurrent neural network. Electron backscattered (EBSD) lab measurements of crystal rotation during rolling were compared with the results of the surrogate model, and despite error introduced by Taylor model simplifying assumptions, very reasonable agreement between the surrogate model and experiment was observed. Our method is foundational for further data-driven studies, enabling the efficient and precise prediction of texture evolution from experimental and simulated crystal plasticity results.

\end{abstract}

\end{frontmatter}

\section{Introduction}
Microstructural science seeks to connect material structure to composition, process history, and properties \cite{molodov2013microstructural}. In crystalline engineering materials, the orientation and distribution of crystals, and their evolution over the course of deformation, are critical microstructural features \cite{bunge2013texture}. Polycrystalline texture impacts several physical \cite{bean1969influence, sauter2014improved}, mechanical \cite{somekawa2017effect}, optical \cite{zaki2018impact, muscarella2019crystal}, chemical \cite{han2018boosting, zhu2020tuning}, and thermal \cite{liu2015anisotropic} properties.

The multitude of crystal plasticity data published in the literature now serves as feedstock for machine learning (ML) studies, in some cases replacing expensive experiments or computationally demanding models \cite{guo2020artificial}. Generally there are two sources of data:

%Electron backscatter diffraction (EBSD) has been the primary tool used to determine the local crystal orientation of crystalline materials since the early '90s \cite{adams1993orientation}. The technique provides quantitative information about texture, and informs on changes during deformation and annealing in metal alloys, mainly via \textit{ex situ} and semi \textit{in situ} experiments \cite{wilkinson2012strains}; data can be collected in 2D or via destructive methods in 3D. New developments at synchrotron X-ray sources have led to breakthrough in non-destructive characterization of crystalline systems, which enable \textit{in situ} characterization of metals in four dimensions (over time in 3D) \cite{renversade2016comparison, shahani2020characterization}. 

i) Experimental measurements: Since the early '90s, electron backscatter diffraction (EBSD) has been the primary tool to generate quantitative information of texture, and track changes during deformation in metal alloys, mainly via \textit{ex situ} and semi \textit{in situ} experiments \cite{adams1993orientation, wilkinson2012strains}. Data can be collected in 2D or via destructive methods in 3D. New developments at synchrotron X-ray sources have led to breakthrough in non-destructive characterization of crystalline systems, which enable \textit{in situ} characterization of metals in four dimensions (over time in 3D) \cite{renversade2016comparison, shahani2020characterization}.

ii) Modelling results: Crystal plasticity models, paired with experiments, track texture evolution during deformation, and do so despite limitations imposed by necessary assumptions. The Taylor model \cite{taylor1938plastic} assumes equality of macroscopic and in-grain plastic deformation rates. Visco-plastic self-consistent (VPSC) model \cite{lebensohn1993self} approximates grains ellipsoidal in interaction with a homogeneous effective medium. ALAMEL \cite{van2005deformation} and GIA \cite{engler2005alloy}, assume velocity gradients is identical to a cluster of grains, rather than individual ones. Crystal plasticity finite element model \cite{roters2010overview}, which is the computationally costly but more accurate. This approach is capable of achieving stress equilibrium at the element boundaries, while also maintaining geometrical consistency throughout the entire representative volume element \cite{roters2011crystal}. Finally, though limited in time and size scale, atomistic simulations address fundamental questions in work hardening mechanisms and crystal rotation during deformation (see e.g. \cite{zepeda2021atomistic}).

Three categories of ML (supervised learning \cite{kotha2019parametrically, salmenjoki2018machine, yang2019predicting, zhao197machine, yang2020exploring}, semi- and un-supervised learning \cite{rovinelli2018using, mansouri2018machine, lei2019machine, liu2020machine} and reinforcement learning \cite{chen2019effective, wang2019meta, mills2020finding}) have been used extensively to predict material properties. For polycrystalline metallic systems, ML has been used to reduce the computational expense of data generation for nonlinear elastic problems \cite{bessa2017framework}, predict fracture patterns in crystalline solids \cite{hsu2020using} and develop constitutive models representing the average properties of the effective medium \cite{mangal2018applied,mangal2019applied}. An entire category of has been efforts devoted to systematic design of a platform to meaningfully prepare the data of polycrystalline samples in 3D and/or 4D \cite{chan2020machine,dai2020graph} .

According to these attempts, regardless of the source of the data, two parameters are essential for a reliable model of plasticity via ML: sufficient data of the representative volume element \cite{bessa2017framework} and a suitable model for handling temporal data \cite{mozaffar2019deep}. However, even when these two parameters are satisfied, ML surrogate crystal plasticity models for spatio-temporal orientation evolution under deformation still report a significant presence of anomalies \cite{shen2019convolutional,pandey2020machine,pandey2020machine2}. Such anomalous behaviour is related to the non-linearity of Euler space, and the presence of multiple equivalent Euler angles in the Euler space due to crystal symmetry \cite{mangal2018applied}. To improve trainability of crystal plasticity data, two solutions have been suggested: 1) defining loss function as averaged disorientation angle to take into account all crystallographic symmetry operators \cite{jha2018extracting}, 2) transferring the data points away from the boundary of the representative space, where identified as prone to low trainability \cite{shen2019convolutional}. These solutions have been partially effective, however, have come at the cost of significant compromise in the computation time. In these cases, the data was sampled from a subdomain of Euler space, whereas, extrapolating forward, one could anticipate that if such a surrogate model is trained to represent the entirety of Euler space, the quantity of outliers should increase, and the trainability should correspondingly decrease.

Here we identify the common source of error, a solution is proposed, and the trained model is tested with experimentally measured data. Crystal orientation evolution data during a rolling process is generated from the entire Euler space using a full-constraint Taylor model (FCTM). The crystal plasticity data first is trained in conventional representations 1) ``as-is'', and 2) after transfer to the fundamental zone. The purpose is to identify the root cause of poor trainability of the data in conventional presentations. After detailing the cause of anomaly a three-step method combining unsupervised machine learning and clustering algorithms is proposed. The method eliminates anomalies and results in the accurate and rapid training of crystal plasticity data. The efficacy of the method is examined using feed forward and recurrent neural networks, and finally, the validity of the orientation evolution modelled by ML, with data from the Taylor model, is assessed by a semi \textit{in situ} crystal rotation measurements during rolling. This new preconditioning approach enables ML-generated high fidelity results and generalization, which opens new avenues in predicting texture evolution.

\section{Nature of the data}
\subsection{Euler angle and Euler space}
\begin{figure}[h]
\centering
\includegraphics[width= 0.45\textwidth]{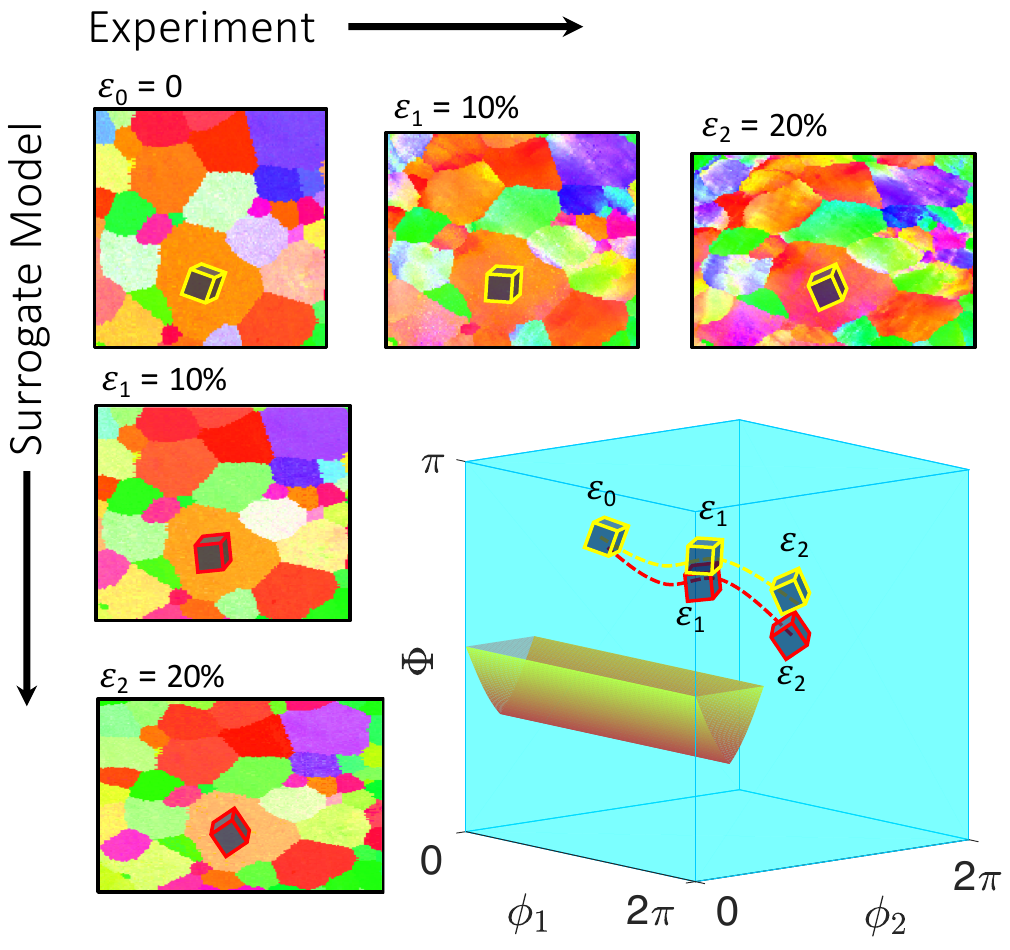}
\caption{\small Any crystal orientation is represented by a coordinate in Euler space, a tetragonal space with $\phi_1$, $\Phi$ and $\phi_2$ axes, coloured in blue. Each crystal orientation with cubic symmetry has 24 equivalent coordinates in Euler space. A change of crystal orientation due to deformation results in a trajectory in this space. The fundamental zone, coloured in copper, is a subset of Euler space in which any orientation appears only once. Our approach is aiming for prediction of this trajectory.}
\label{Fig:EulerSpace}
\end{figure}

Crystallographic orientation is described by three Euler angles ($\phi_1$, $\Phi$, $\phi_2$), that represent the relation between the axes of the crystal lattice with respect to a fixed coordinate system. For the rolling process, the reference coordinate system is the rolling, transverse and normal directions (RD, TD and ND, respectively). By employing the three Euler angles, an orientation can be represented by a coordinate in the Euler space with perpendicular axes of length $2\pi$, $\pi$ and $2\pi$ for $\phi_1$, $\Phi$ and $\phi_2$, respectively. Lattice deformation results in rotation of the crystal, which is equivalent to the change of coordinate in the Euler space.  As shown schematically in Figure \ref{Fig:EulerSpace}, crystal orientation of any scanned point is mapped to a coordinate in the Euler space, thus, during deformation a trajectory is described. This work proposes preconditioning the data such that the surrogate model for trajectory prediction represents the minimum deviation from the experimentally measured values, no matter the initial data's origin in Euler space.

Depending on the stacking sequence and arrangement of atoms, a crystal lattice will show certain elements of symmetry. The presence of crystal symmetry will result in a different but equivalent number of points in the Euler space, which all represent the same orientation. For instance, Aluminum (used in this work), with face centred cubic lattice, is invariant under 24 different symmetry operators; thus, for any given orientation, 24 equivalent points appear in the Euler space. The so-called fundamental zone (FZ) of the Euler space is the subset of Euler space (or orientation space in general) within which each orientation appears only once. The FZ of the Euler space is defined in the range of $\left[0, 2\pi\right]$, $\left[0, \pi/2\right]$ for $\phi_1$ and $\phi_2$, respectively.  $\Phi$ varies in the range of $\left[0, \pi/2\right]$, and $\cos \Phi>\frac{\cos \phi_2}{\sqrt{1+\cos^2\phi_2}}$ and $\cos \Phi>\frac{\cos \left(\frac{\pi}{2}-\phi_2\right)}{\sqrt{1+\cos^2\left(\frac{\pi}{2}-\phi_2\right)}}$ are also satisfied. The Euler space, axis and FZ, are shown in Figure \ref{Fig:EulerSpace}. A comprehensive description of the orientation representation, Euler angles, Euler space, FZ, the method of calculation and the corresponding Python code are available in the Supplementary Information (SI).

Our ML algorithms are trained on the dataset generated by the fully constrained Taylor model (FCTM) \cite{bunge1970some}. FCTM is perhaps one of the simplest computational approaches to predicting plasticity in metals.  It has relatively low computational overhead and yet has been shown to be able to capture many features of e.g., texture evolution in metals \cite{kocks2000texture}.  FCTM neglects the interactive/constraining effects of neighbouring grains on the path of deformation. This assumption imposes error on the prediction of texture evolution in real materials; however, this method is an ideal candidate for data generation for the purpose of testing the trainability of the crystal plasticity data due to three reasons: (i) it is possible to sample the entire Euler space, which is not easily achievable by other models, (ii) the only input of the Taylor model is the initial orientation of the crystal and the mode of deformation, thus, there is no size limitation for the generated dataset, and (iii) the simplification assumption of the Taylor model eliminates the features of the training set which are related to interactions between neighbouring crystals. Thus, the intrinsic error associated with the representation of the data in Euler space will be measurable. When generating the training and test set, we discretized the Euler space in increments of 8 degrees, wherein each initial orientation was deformed up to 30\% with increments of 1\%, for a total of 1,336,500 data points. 

Two metrics are used to compare model prediction and the ground truth: 1) the absolute difference between the predicted Euler angles and the ground truth ($\Delta \varphi$), and 2) disorientation angle, defined as the smallest possible rotation angle among all symmetrically equivalent misorientations.

\section{Anomaly in training of CP data}\label{sec:ANN}
Initially, two conventional formats of CP data representation are trained: 1) Orientation evolution obtained from FCTM, as-is. In this case, the initial orientation is uniformly distributed in the Euler space and the subsequent orientation change due to deformation is tracked. 2) Prior to training, the data is transferred to the FZ. The results are used to identify the shared features of outliers reported previously \cite{shen2019convolutional,pandey2020machine,pandey2020machine2}, and used to inform the solution accordingly.  

%Three cases are considered to examine the trainability of CP data: 1) in which the data is trained as-is, 2) prior to training, the data is transferred to the FZ, and 3) local density-based clustering algorithm, DBSCAN, is included to improve the trainability of the data.  The details of the preconditioning approaches for Cases 2 and 3 are discussed in sections \ref{Sec:FZ} and \ref{Sec:Clustered}, respectively. We also examine the effect of the number of hidden layers on the trainability of the data. In each case, 60\%, 20\% and 20\% of the data is used in training, validation and test sets, respectively, unless stated otherwise.  Scenarios considered in this work to evaluate trainability of the CP dataset are summarized in Figure \ref{Fig:flowchart}. 

%\begin{figure}[h]
%\centering
%\includegraphics[width= 0.4\textwidth]{flowchart.pdf}
%\caption{\small Flow chart showing the sequence of preconditioning steps and architecture of ML models. Case 1 trains the data as-is from Taylor model. Case 2 transfers the CP data to FZ. Case 3 preconditions the data using a suggested method of clustering. The data is then trained by two ANN models with 2 and 5 hidden layers and 100 neurons for each layer. Case 4 data is trained using SLTM and GRU. The weights calculated in Case 3 are then used to predict the texture evolution in the experimental data.}
%\label{Fig:flowchart}
%\end{figure} 

\subsection{Outliers for uniformly distributed data in Euler space}\label{Sec:asis}

\begin{figure*}[h]
\centering
\includegraphics[width= 1\textwidth]{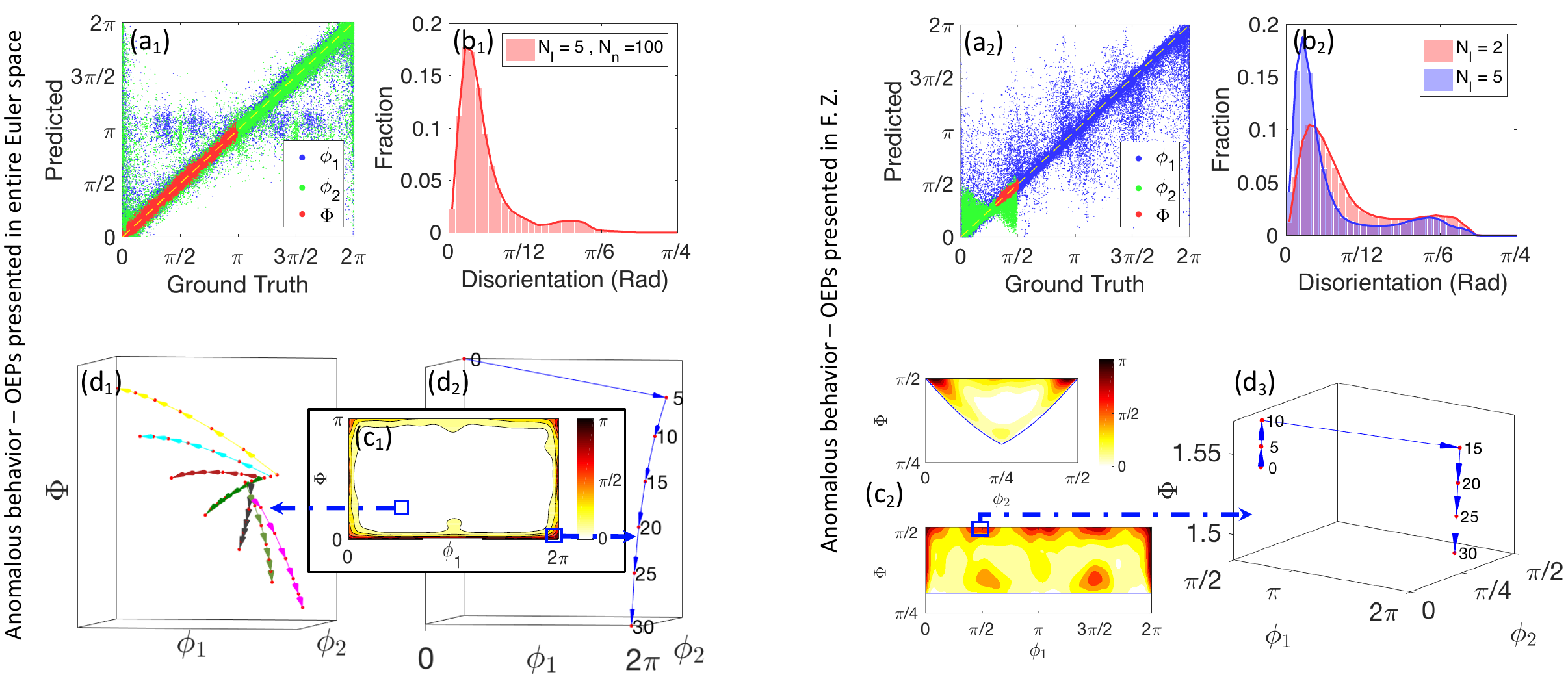}
\caption{\small  Illustrating the anomalous response of crystal plasticity data to training in conventional representations. a$_1$ and a$_2$) Comparison between artificial neural network prediction and the measured values from fully constrained Taylor model (ground truth) for orientation evolution path initiated in Euler space (uniformly) and after transformation to the fundamental zone, respectively. b$_1$ and b$_2$) The distribution of disorientation angle between the measured and predicted values. Two peaks observed in each representation indicate that for both cases the trainability falls into two regimes. Increasing the number of artificial neural network hidden layers from $N_l = 2$ to $N_l = 5$ improves the prediction slightly; however, the model still fails in reliably predicting the texture evolution. c$_1$ and c$_2$) $\Delta \varphi$, which is noticeably larger when the initial crystal orientation is adjacent to the boundary compared to those away from the boundary of the Euler space or the fundamental zone. d$_1$) Typical orientation evolution path far from the boundary of Euler space.  d$_2$ and d$_3$) orientation evolution path near the boundary of Euler space and the fundamental zone, respectively, in which the path is disconnected due to exiting the corresponding space.}
\label{Fig:ESFZ}
\end{figure*}

The architecture of the feed forward artificial neural network (ANN) and the models predictions are shown in Figure S6(a). The input layer includes 4 features of initial Euler angles ($\phi_1^i$, $\Phi^i$, $\phi_2^i$) and the applied reduction ($\varepsilon$) via rolling process. The output layer consists of Euler angles after plane strain compression mode of deformation. Figure \ref{Fig:ESFZ}(a$_1$) shows the comparison between the predicted Euler angle of the test set at the end of the training process, and the ground truth (FCTM). $\phi_1$, $\Phi$ and $\phi_2$ are coloured in blue, red and green respectively. There are significant number of outliers, mainly at $\phi_1 =\pi$ and $\phi_2 = 0 $ and $2\pi$. The disorientation distribution and absolute difference between the ground truth and predicted Euler angles are presented in Figure \ref{Fig:ESFZ}(b$_1$) and (c$_1$), respectively. For ANN with $N_l = 5$ and $N_n = 100$, disorientations are as high as $\pi/5$ and form almost two distinct peaks, implying two regimes in the data. This effect is also observed in $\Delta \varphi$ map in the $\phi_1-\Phi$ projection of Euler space. While the data far from the boundaries of the Euler space shows reasonable trainability, the error increases in the vicinity of the boundary. Let us define \textit{Orientation evolution path (OEP)}, as the curve resulting from connecting the coordinates in the Euler space from the least to the most reduction value (0 to 30\% in this case). Figure \ref{Fig:ESFZ}(d$_1$) displays 7 OEPs, whose initial orientations alter along $\phi_2$ in increments of 8 degrees. Overall, the path is smooth and the associated error is reasonably small. The major error, however, results from cases such as the one shown in Figure \ref{Fig:ESFZ}(d$_2$), with discontinuity in EOP due to exiting the Euler space and re-entering from a different region.

%\begin{figure}[h]
%\centering
%\includegraphics[width= 0.45\textwidth]{ES.pdf}
%\caption{\small  Anomalouda) Comparison between the prediction of the model and the ground truth (FCTM). b) The distribution of disorientation angle between the measured and predicted values. Two peaks indicate that the data generated by discretization fall into two regimes in terms of trainability.  c) $\Delta \varphi$, which is noticeably larger when the initial crystal orientation was adjacent to the boundary compared to those away from the boundary of the Euler space. d) Typical OEP far from the boundary and e) OEP near the boundary, in which the path is disconnected due to exiting the Euler space.  }
%\label{Fig:ES}
%\end{figure} 

\subsection{Outliers after transferring data to fundamental zone}\label{Sec:FZ}
In FCC systems, crystal symmetry results in 24 dissimilar but equivalent OEPs in Euler space, such as those shown in Figure S7. Presenting the Euler angles that lie within the FZ is an approach to avoid the repetition of equivalent data \cite{demir2009taylor, pandey2020machine}.

%Any given crystal orientation is represented by 24 equivalent but unequal coordinates in Euler space. Therefore, there are 24 equivalent OEP, represent the rotation of the crystal over the course of deformation. An example of that is shown in Figure \ref{Fig:AllClusters}. A FCC crystal with initial orientation of $(0\ ,\ 0.97\ ,\pi/4)$ is deformed by plane strain mode of deformation, thus, the orientation alters and forms a path. However, 24 equivalent paths forms due to the symmetry of cubic crystal structure. These paths are shown with red arrows and the the arrows are in the direction of increase of strain. To avoid repetition of dissimilar but equivalent OEPs, one way is to chose the data from FZ, shown in blue in Figure \ref{Fig:AllClusters} \cite{demir2009taylor, pandey2020machine}. 

Treatment of data by transferring to the FZ, Figure S8(b) shows the distribution of the Euler angles. Due to the irregular geometry of the FZ, the data distribution is non-uniform. A comparison between the predicted Euler angles and the ground truth (calculated by FCTM) for the test set is shown in Figure \ref{Fig:ESFZ}(a$_2$). Even though many data points lie on the centre line, significant outliers are scattered, indicating poor prediction of the model, mainly for $\phi_1$ and $\phi_2$. 
The distribution of the disorientation angle between measured and predicted Euler angles is shown in Figure \ref{Fig:ESFZ}(b$_2$). Although increasing the number of NN hidden layers shifts the disorientation curve to smaller values, it is not sufficient to serve as a reliable surrogate model.

%\begin{figure}[h]
%\centering
%\includegraphics[width= 0.45\textwidth]{FZ.pdf}
%\caption{\small (a) Comparison between ANN prediction and the measured values from FCTM (ground truth). Significant deviation from the centre line is observed and implies failure of the ML model in predicting texture evolution. (b) Disorientation angle, as a metric of error, for  $N_l = 2$ and $N_l = 5$. Increasing the number of NN hidden layers improves the prediction slightly; however, the model still fails in reliably predicting the texture evolution. (c) Map of $\Delta \varphi$, with hotspots at the border of the FZ and multiples of $\phi/2$. (d) An example of OEP which intersects the boundary of the FZ. Predictability of such paths is low. }
%\label{Fig:FZ}
%\end{figure}   

Figure \ref{Fig:ESFZ}(c$_2$) shows the projection of $\Delta \varphi$ on $\phi_2-\Phi$ and   $\phi_1-\Phi$ planes, respectively. In a loose generalization, hotspots of the error map are concentrated in two parts: adjacent to the boundary of the FZ, and at the multiples of $\pi/2$. Considering the non-linear nature of the crystal plasticity problems, curvature, direction, smoothness and uniformity of OEPs is inconsistent (12 examples of OEPs are shown in the Figure S9). However, the main source of error is discontinuity of OEP due to exit from the FZ and entry from another location. An example of such OEP is displayed in Figure \ref{Fig:ESFZ}(d$_3$). The initial Euler angles are $(1.68,1.54,0.70)$. Deformation results in departure of FZ and a sudden change of direction and corresponding jump in the OEP. Such data points have low trainability.

\section{OEP preconditioning via discontinuity removal}\label{Sec:Clustered}
\begin{figure*}[h]
\centering
\includegraphics[width=1\textwidth]{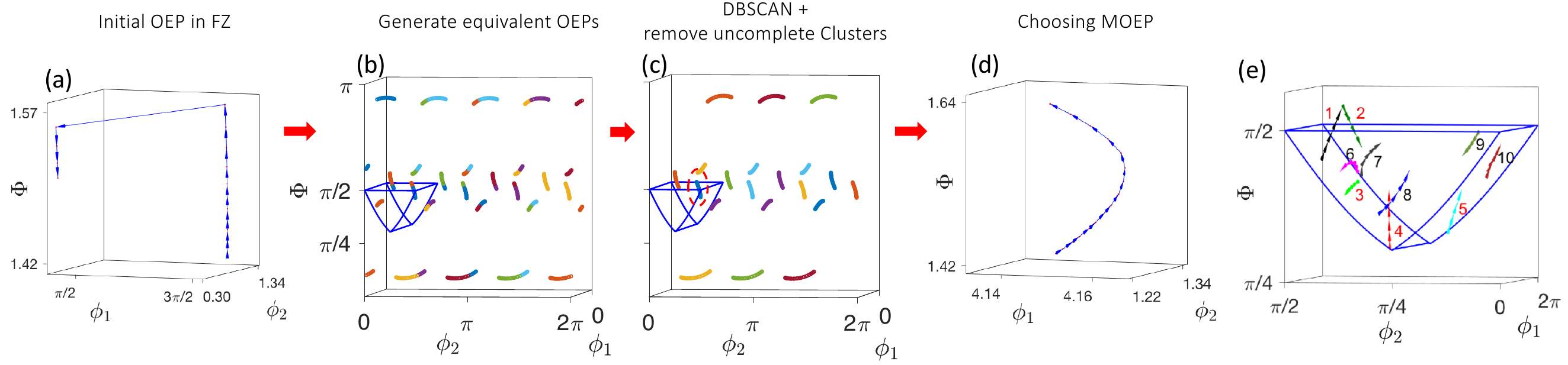}
\caption{\small Steps of OEP modification: (a) for any given OEP in the FZ, there is a possibility of abrupt orientation change. (b) 24 equivalent OEPs are generated. (c) Data points are clustered using a density based clustering algorithm and clusters of the size 31 with non-repeating labels strain are saved. If more than one cluster satisfies the criteria, that with the maximum data points in the FZ is chosen.  (d) The abrupt change in the direction and length of OEP due to the geometrical restriction of the FZ is removed.  e) Examples of OEP after applying DBSCAN clustering algorithm, which eliminates the data point discontinuity. The modified orientation evolution path is referred to as MOEP. Paths 1-5 are examples of the modified paths, while paths 6 - 10, which are are fully within the FZ, remain the same.}
\label{Fig:DBSCAN}
\end{figure*}

Having identified discontinuity of OEP as the cause of poor data trainability, here we suggest a data preconditioning procedure. The procedure includes three major steps that are shown in Figure \ref{Fig:DBSCAN}, and the corresponding codes are available in the SI. 

%Step  1: Generate the equivalent orientations: Any given OEP, consist of 31 orientation matrixes ($g^{j}$), where $j$ is the deformation increments varies in the range of [0,30]. An example of that is shown in Figure \ref{Fig:DBSCAN}(a). FCC crystals with cubic symmetry possess 24 symmetry operators ($O_{i}$), where $ 1 \leq i \leq 24$. Thus, $g_i^j = O_{i} g^j $, including 24$\times 31$ orientations matrixes are generated, shown in Figure \ref{Fig:DBSCAN}(b). The points that generated in the same iteration over  $O_{i}$ are colored the same. 

Step  1: Generate the equivalent OEPs. Any given OEP, consisting of 31 orientation matrices ($g^{j}$), where $j$ is the deformation increments and varies in the range of [0,30]. An example is shown in Figure \ref{Fig:DBSCAN}(a). FCC crystals with cubic symmetry possess 24 symmetry operators ($O_{i}$), where $ 1 \leq i \leq 24$. Thus, $g_i^j = O_{i} g^j $, including 24$\times 31$ orientation. Figure \ref{Fig:DBSCAN}(b) shows $g_i^j$, in which orientations with equal $i$ share the same colour.

Step 2: Cluster the orientations. Data points labeled by the deformation increment in the range of (0-30) are clustered. The total number of clusters $N_c \geq 24$, depends on the initial crystal orientation. 24 clusters form when none of the OEPs intersect with the boundary of the Euler space, whereas when intersection occurs, the number of clusters increases. Local density-based clustering algorithm \cite{ester1996density}, DBSCAN, is the method be suited to clustering such data, since there is no need to specify the number of clusters, and it also excels at identifying clusters of non-spherical shapes. Only clusters of size 31 with unique labels of 0 to 30 are selected, as shown and coloured in Figure \ref{Fig:DBSCAN}(c). 

Step 3: Select the cluster. If multiple clusters satisfy this criteria, the one with the highest number of data points inside the FZ is selected. One such cluster is highlighted and magnified in Figure \ref{Fig:DBSCAN}(c) and (d). Although OEPs in Figure \ref{Fig:DBSCAN}(a) and (d) are equivalent, the latter does not exhibit a discontinuity due to constraints of the representative space including the Euler space and FZ. Hereafter, the modified OEP is referred to as MOEP and the combination of the MOEP and the ANN is referred to as MOEP-ANN.

Figure \ref{Fig:DBSCAN}(e) shows the number of MOEPs and their relative position with respect to FZ. For enhanced visualization, only data points for which $\varepsilon$ is a multiple of 6 are plotted.  
Comparing the data before and after modification, this procedure only modifies the OEPs which cross the boundary of the FZ; those which are fully embedded in the FZ are identical in both cases. Paths 1-5 are examples of the former, while paths labeled 6 - 10 in Figure \ref{Fig:DBSCAN}(e) are examples of the latter. 

%\begin{figure}[h]
%\centering
%\includegraphics[width= 0.35\textwidth]{Clustered.pdf}
%\caption{\small Examples of OEP after applying DBSCAN clustering algorithm, which eliminates the data point discontinuity. Paths 1-5 are examples of the modified paths, while paths 6 - 10, which are are fully within the FZ, remain the same.}
%\label{Fig:Clustered}
%\end{figure} 

\begin{figure*}[h]
\centering
\includegraphics[width=0.9\textwidth]{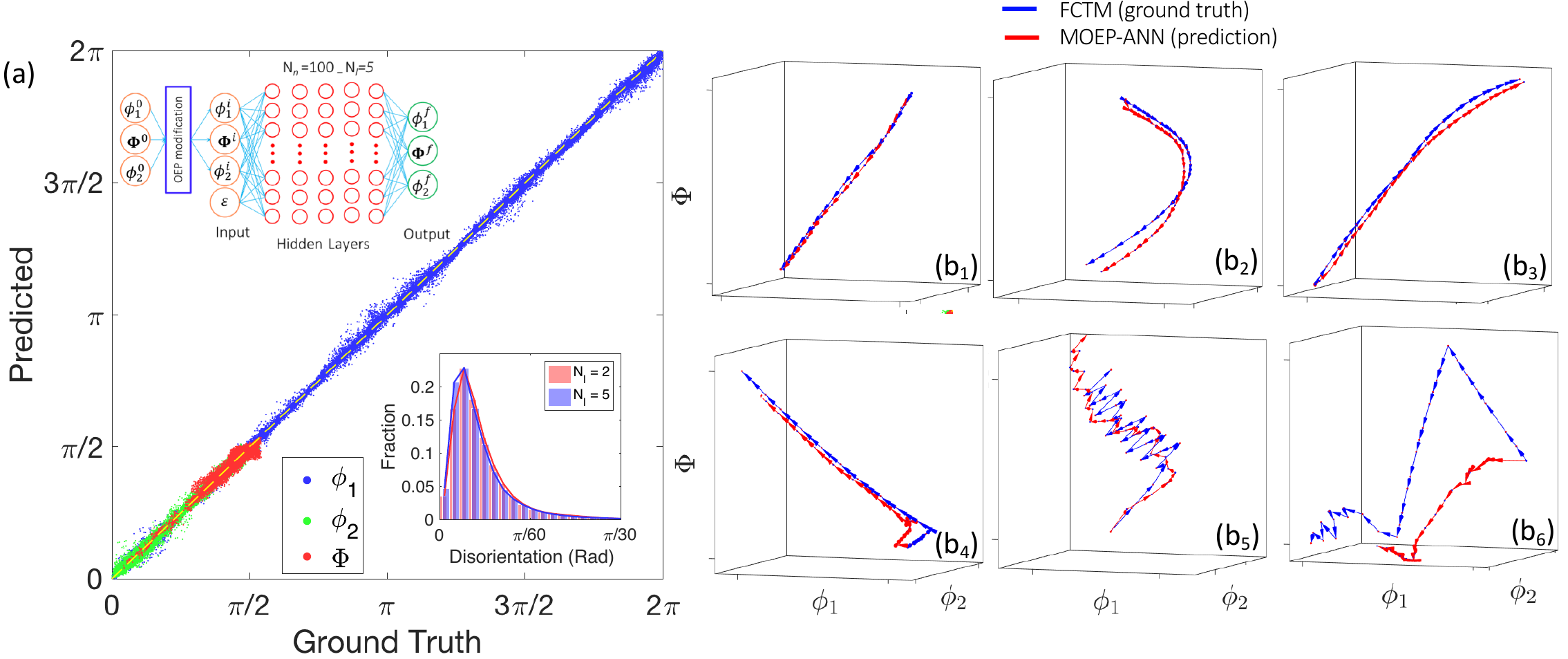}
\caption{\small (a) The MOEP-ANN predicted Euler angles versus the ground truth (FCTM) for the test dataset, also, Graph of disorientation distribution between MOEP-ANN and FCTM, for both 2 and 5 hidden layers and 100 neurons. The peak of the disorientation curve is at 0.8 degrees for both neural network architectures. (b) Comparison between the predicted and ground truth OEPs. Two regimes of OEPs are observed: 1) (b$_1$-b$_3$) are examples of smooth paths, for which, prediction of the model is in very good agreement with the ground truth. Over 90\% of the OEPs are from this type. 2) (b$_4$-b$_6$) are the other regime of OEPs, which include sharp change of directions. The abrupt change can be attributed to the switch between different slip systems or the Taylor model ambiguity. }
\label{Fig:ClusteredResults}
\end{figure*}

The effect of the precondition procedure on the trainability of the data is presented in Figure \ref{Fig:ClusteredResults}. Very good agreement between the predicted Euler angles for the test set vs. the ground truth is observed in Figure \ref{Fig:ClusteredResults}(a). No significant outliners and very limited deviation from the centre line indicates the effectiveness of the preconditioning algorithm. The distribution of the disorientation angle for models with 2 and 5 hidden layers is embedded in Figure \ref{Fig:ClusteredResults}(a). The peak in error distribution occurs at 0.8 degrees, with 95\% of disorientation angles less than 3$^\circ$. Figure \ref{Fig:ClusteredResults}(b) is the direct comparison between the trajectory of orientation evolution from the Taylor model (blue arrows) and MOEP-ANN (red arrows). A group of trajectories, such as that in Figure \ref{Fig:ClusteredResults}  (b$_1$-b$_3$) form a smooth path, and the model has a very good agreement with the ground truth. Over 90\% of the MOEPs are from this type. There are, however, some MOEPs for which sudden changes in the trajectory are observed. Such sharp changes in direction can be attributed to activation of a different set of slip systems. Such abrupt change was successfully captured in the cases of Figure \ref{Fig:ClusteredResults}(b$_4$), while the model could not capture the essence of the ground truth in case Figure \ref{Fig:ClusteredResults}(b$_6$). While this can be a normal response of a crystalline system to deformation, in some cases another possibility explains the irregular MOEPs: Taylor ambiguity. Taylor model searches for a set of five independent slip systems which minimize the sum of the internal work done to accommodate the imposed deformation. The set of activated slip systems determines the lattice rotation during plastic deformation and eventually the direction of the MOEPs. However, a problem arises when there is more than one combination of five independent slip systems with the same minimum of total shear \cite{manik2014review,holmedal2020regularized}. In this case, each combination will result in a different lattice rotation and the model might have chosen either of two sets of slip systems. Unexpected changes in direction or serrated MOEPs, such as the one shown in Figure (b$_5$), might be due to this effect. Considering the correlation between the error and sudden change of direction, MOEP-ANN can be used to determine the possible change of slip systems.

\section{Recurrent neural network results}
In section \ref{sec:ANN} the input layer consists of the crystal orientation before deformation and the applied strain. An alternative approach to modeling the CP data is fitting a RNN to learn from the crystal orientation of the previous increments to predict the orientation in the current increment. Here we employed GRU, a gating mechanism in recurrent neural networks, to describe crystal orientation as a map that depends on the orientation history of previous increments in deformation. Trajectories of lengths 6 and 9 with increments of 1\% reduction served as model inputs, and GRUs then compute the crystal orientation in the next immediate strain increment. The steps of constructing the sequential data is available in the Figure S10. The GRU architecture is shown in Figure S6(d). Prior to training the sequential data, the preconditioning process presented in section \ref{Sec:Clustered}, is applied. GRUs use history-dependent hidden states $h\left(\varepsilon_{j-1}\right)$ to compute the outputs $\phi_1^o(\varepsilon_j)$, $\Phi^o(\varepsilon_j)$, $\phi_2^o(\varepsilon_j)$.  

\begin{figure}[h]
\centering
\includegraphics[width=0.4\textwidth]{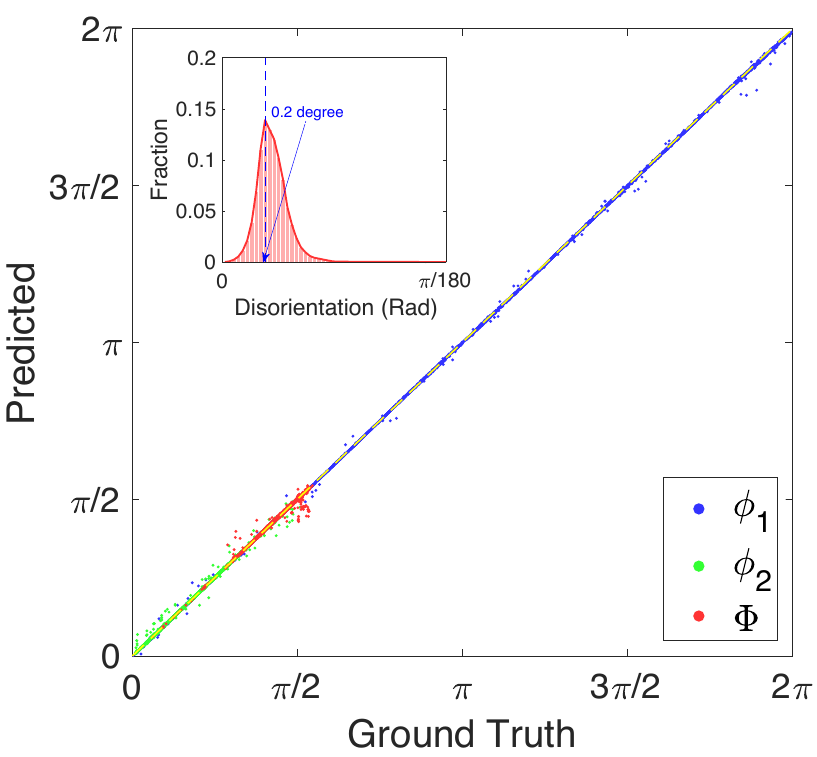}
\caption{\small (a) The MOEP-RNN predicted Euler angles versus the ground truth and disorientation distribution between them, shows the success of the surrogate model in predicting the orientation evolution during the rolling process trained by Taylor model data. The structure of GRU network in this work, including preconditioning process, input layer, a matrix of 3$\times 5$ and output of the size of 3$\times 1$ for each sample.}
\label{Fig:RNN4}
\end{figure}

The comparison Figure \ref{Fig:RNN4}, shows excellent agreement between the predicted orientation and the ground truth for trajectories of the length of 5 strain increments. The disorientation distribution between the GRU model prediction and the FCTM is embedded in Figure \ref{Fig:RNN4}. The mode of distribution is 0.2$^\circ$, indicating the GRU model employed in this work is able to predict the spatially and temporally resolved 3D microstructure evolution under plane strain compression loading (rolling process).

%\begin{figure}[h]
%\centering
%\includegraphics[width= 0.45\textwidth]{loss.pdf}
%\caption{\small }
%\label{Fig:loss}
%\end{figure} 

\section{Prediction of rolling texture evolution from ML results}
To further examine the surrogate model trained by FCTM data, the performance was tested on experimentally measured texture evolution data. Fully recrystallized commercially pure aluminum sheet was rolled at room temperature in four consecutive steps to apply $\varepsilon = 6\%,\ 12\%,\ 18\%$ and 24\% reduction. Using EBSD analysis, local crystallographic orientations were measured at each rolling step. To ensure that the same group of grains from the bulk were tracked throughout the series of deformation, a semi \textit{in situ} technique known as split-sample was conducted. The sample preparation details and EBSD measurements are presented in \cite{pirgazi2020semi}.

\begin{figure}[h]
\centering
\includegraphics[width=0.4\textwidth]{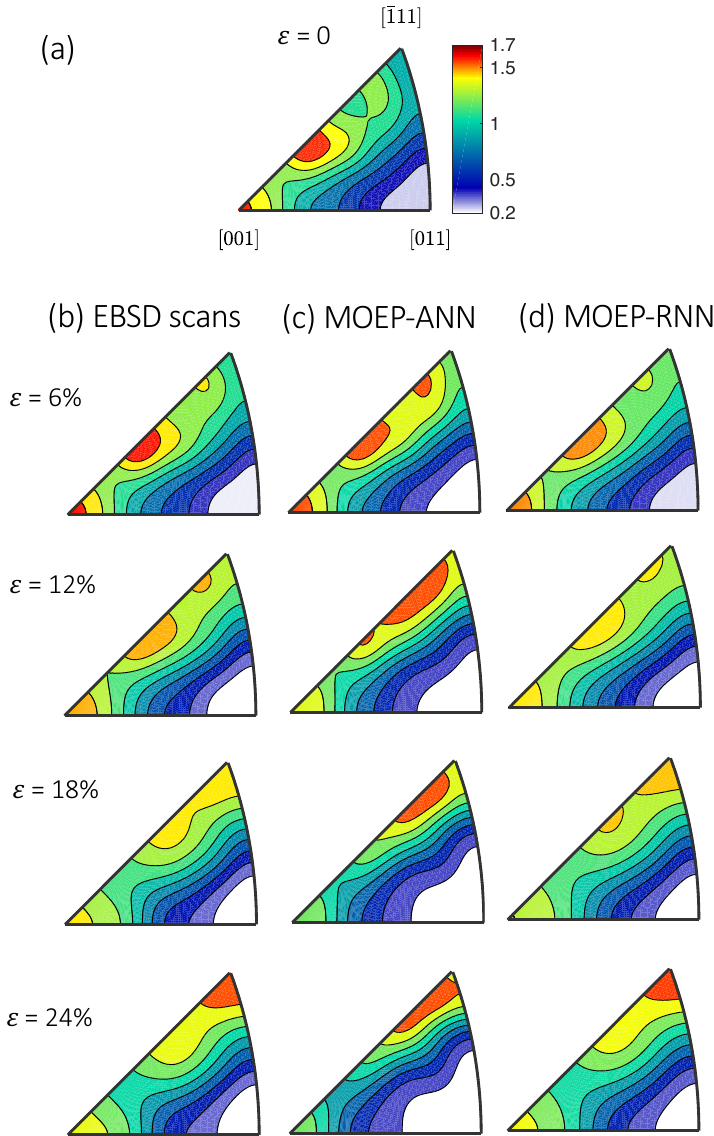}
\caption{\small Comparison between the experimentally measured crystallographic orientation evolution and predication of the ML models. (a) The RD-IPF contour of the annealed sample. (b) The RD-IPF contour of the annealed sample after applying $\varepsilon = 6\%,\ 12\%,\ 18\%$ and $24\%$ reduction via rolling. (c) is the equivalent set of contours predicted by MOEP-ANN. (d) is the equivalent set of contours predicted by MOEP-RNN. A good agreement between the experimental measurements and prediction of surrogate model is observed, most notably, for our MOEP-RNN model.}
\label{Fig:IPFDistribution}
\end{figure}

Euler angles at each EBSD scan point (step size of $2 \mu m$ $ \approx$ 590,000 data points) were extracted and the evolution was tracked over the course of deformation. Figure \ref{Fig:IPFDistribution} (a) and (b) shows the RD-inverse pole Figure (RD-IPF) contours of the samples in annealed state, and after subsequent deformation. Predication of MOEP-ANN and MOEP-RNN are presented in Figure \ref{Fig:IPFDistribution} (c) and (d) respectively. While MOEP-ANN shows good prediction of the experiment up to $\varepsilon = 12\%$, the contours resulting from the MOEP-RNN are in excellent agreement with EBSD measurements for the entire range of deformation.  An alternative presentation of crystallographic orientations is through IPF colour assignment. Figure S11 (b$_1$-b$_5$) illustrates the pixel-by-pixel prediction of the orientation evolution for MOEP-ANN, wherein Figure S11 d$_1$ shows the distribution in disorientation angle between all grain pairs (predicted by MOEP-ANN model and experiment). Where the deformation is below 15\%, the average disorientation angle for all orientations is below $\pi/18$, and by gradually increasing the deformation the curves shift to larger values. This degree of error is exactly in the range of intrinsic error of the Taylor model \cite{pirgazi2020semi}. Since in the best case scenario the surrogate model is as accurate as the ground truth (in this case FCTM), any improvement will come from training the model with more accurate data.

The more accurate prediction in MOEP-RNN compared to MOEP-ANN is partially due to higher model accuracy; however, the primary source of improvement is due to the relatively short input trajectory length in the MOEP-RNN, at an interval of 6 yielding an intrinsic error of FCTM of up to 6\%. In MOEP-ANN, by contrast, the trajectory may deviate from the experiment by up to 24\%. A detailed discussion on the source of error of the two models is available in SI 12. 

The results show that crystal plasticity temporal data exhibit excellent trainability, with removal of the source of anomaly. A summary of the architecture performance differences and the effect of the preconditioning procedure used in this work are shown in Table S1. Two major benefits in our preconditioning approach are identified: 1) the test score is significantly enhanced: from 0.83 to to nearly unity (perfect), and 2) the convergence condition is met after only 16 iterations, while without preconditioning the convergence condition is not satisfied even with a magnitude more iterations.

%\begin{figure*}[h]
%\centering
%\includegraphics[width=1\textwidth]{EBSD.pdf}
%\caption{\small Comparison between the experimentally measured crystallographic orientation evolution and predication of the ML models. (a$_1$-a$_5$) is the ND-inverse pole figure maps with $\varepsilon = 6\%,\ 12\%,\ 18\%$ and $24\%$. (b$_1$-b$_5$) is the equivalent set of micrographs predicted by MOEP-ANN.  (c$_1$-c$_5$) is the equivalent set of micrographs predicted by MOEP-RNN. The model applied to all pixels of the micrograph. A good agreement between the experimental measurements and prediction of surrogate model is observed, most notably, for our MOEP-RNN model. (d$_1$ and d$_2$) are the disorientation distribution between model prediction and EBSD results in a grain-by-grain comparison for MOEP-ANN and MOEP-RNN, respectively.}
%\label{Fig:EBSD}
%\end{figure*} 

\section{Methods}
\subsection{Feed-forward neural network}
We trained ANN with 4 features in the input layer including three Euler angles, representative of the crystal orientation before deformation ($\phi_1^i$, $\Phi^i$, $\phi_2^i$), and the applied reduction ($\varepsilon$). The output layer has three neurons that are the Euler angles after deformation via rolling process ($\phi_1^f$, $\Phi^f$, $\phi_2^f$). $N_l$ hidden layers, with equal numbers of neurons, $N_n$, are designed. $N_l$ and $N_n$ vary and are specified for each case. Rectified Linear Unit (ReLU) and a stochastic gradient-based optimizer (adam) were used as the activation function and the solver for weight optimization, respectively. In each case, 60\%, 20\% and 20\% of the data is used in training, validation and test sets, respectively, unless stated otherwise.

\subsection{Gated recurrent neural network}
RNN are a class of neural networks designed to learn from sequential events. We used a specific subset of RNN, gated recurrent unit (GRU) \cite{cho2014learning}. The advantage of GRU is in its capability to avoid the vanishing gradients phenomenon by using multiple data-gate mechanisms that control the flow of storing or forgetting information in hidden states and outputs. Compared to Long Short-Term Memory (LSTM), GRU's formulation is less prone to overfitting. 

The CP data of this work, in the sequential form, is a trajectory including Euler angles $\phi_1^i(\varepsilon_j)$, $\Phi^i(\varepsilon_j)$, $\phi_2^i(\varepsilon_j)$, where $0 \leq j \leq 30$ is the incremental deformation. Each trajectory is split into $t+1$ steps, where $t$ is the length of history and the Euler angles at increment $t+1$ is the ground truth or the output. For instance, at $t = 9$, Euler angles at $[\varepsilon_0,\varepsilon_8]$,  $[\varepsilon_{10},\varepsilon_{18}]$, $[\varepsilon_{20},\varepsilon_{28}]$ are the input array and $[\varepsilon_9]$, $[\varepsilon_{19}]$, $[\varepsilon_{29}]$ are the ground truth or output. The data point at $[\varepsilon_{30}]$ is ignored. In this work, $t = 6$ is considered. Input data is normalized to a range between 0 and 1. 

\subsection{Fully Constrained Taylor Model}
For any given crystal orientation, FCTM determines the texture evolution by searching for a set of five independent slip systems (out of 12 possible systems in FCC alloys) which minimize the sum of the internal work done in the grain, and which accommodate the imposed deformation. The main assumption of this model is that, in a polycrystalline system, the deformation of all grains is identical to the macroscopic deformation. We used the MTM-FHM software \cite{Houtte1995} for the FCTM simulations of deformed textures with plane strain compression as the imposed mode of deformation.

$\Delta \varphi$ as one of the error metrics is calculated by discretizing the  domain to the cubic volume elements of the side size of $\pi/25$ and the error is averaged in each grid. The discrete values of error are then interpolated by gaussian process regression \cite{rasmussen2003gaussian}. The details of disorientation angle calculations are available in the SI.

\section{Conclusion}
We have developed a preconditioning approach which overcomes a fundamental problem in learning from crystal plasticity data which arises due to the strong influence of data anomalies. The anomaly source was identified by applying ANN to two datasets: (i) Sampling from the entire Euler space by equally discretizing the Euler space, followed by 30\% deformation. (ii) Transferring the orientations to the fundamental zone as a subspace with a single representation of any given orientation. Such formats are conventional representations of crystal plasticity data; however, these models exhibit poor performance in the vicinity of the spaces boundary. The three step unsupervised preconditioning approach then used to modify the orientation evolution path by removing the anomalies for those close to the boundary of the space, and enhanced trainability dramatically. 

The efficacy of this approach was also examined by RNN. The mean disorientation between ground truth and predicted value is smaller than 0.2$^\circ$, indicating that the proposed surrogate model serves as an effective substitute for crystal plasticity calculations. The proposed algorithm is applicable to any mode of deformation. Training the data under any loading condition enables obtaining a coefficient which can be used to reliably model the crystallographic orientation evolution for the corresponding mode of deformation. Another advantage of this method is that the training set has not been limited to a local region of the Euler space, but rather covers the entire Euler space.

Finally, the validity of the deformation texture calculated from the surrogate model is assessed by a semi \textit{in situ}  measurements of crystal rotation during rolling. Generally, the model is in good agreement with the experimentally measured texture.

\section{Supporting Information} 
The following file is available free of charge.
\begin{itemize}
  \item SI.pdf: Includes representation of Euler angle and space, fundamental zone, misorientation, IPF color assignment, curves of loss functions and more examples of OEPs. 
  
    \item SI.files: Includes python codes for calculation of disorientation angles, IPF colours, transfer to FZ and training the data sets for ANN and GRU. Matlab codes for procedure of OEP modification is also available. http://clean.energyscience.ca/codes

\end{itemize}

\section{Acknowledgement} 
The financial support received from NSERC / UNENE Research Chair in Nuclear Materials is greatly acknowledged. I.T. acknowledges NSERC and performed work at the NRC under the auspices of the AI4D and MCF Programs.

\pagebreak
\small
\section{References}
\bibliographystyle{ieeetr}
\addcontentsline{toc}{section}{References}
\bibliography{Biblio_NN_Taylor.bib}

\end{document}